\documentclass[journal=jacsat,manuscript=article]{achemso}

\usepackage{bm}
\usepackage{graphicx}
\usepackage{overpic}
\usepackage{latexsym}
\usepackage{amsmath}
\usepackage{amsfonts}
\usepackage{amssymb}
\setkeys{acs}{articletitle = true}
\SectionNumbersOn

\newcommand{\TC} {\mbox{\scriptsize TC}}

\newcommand{\eff} {\mbox{\scriptsize eff}}

\def \MMO  {{M^{-1}}}
\def \MMF  {{M^{-5/3}}}

\title{Combining the Transcorrelated method with Full Configuration Interaction Quantum Monte Carlo: application to the homogeneous electron gas}

\author{Hongjun Luo}
\email{h.luo@fkf.mpg.de}
\affiliation{Max-Planck-Institut for Solid State Research, Heisenbergstra\ss e 1, 70569 Stuttgart, Germany}
\author{Ali Alavi}
\email{a.alavi@fkf.mpg.de}
\affiliation{Max-Planck-Institut for Solid State Research, Heisenbergstra\ss e 1, 70569 Stuttgart, Germany}
\alsoaffiliation{Department of Chemistry, University of Cambridge, Lensfield Road, Cambridge, CB2 1EW, United Kingdom}

\begin{document}

\begin{abstract}
\noindent  We suggest an efficient method to resolve electronic cusps in electronic structure calculations, through the use of an  
effective transcorrelated Hamiltonian. This effective Hamiltonian takes a simple form for plane wave bases, containing up to two-body operators only, and its use incurs almost no additional computational overhead compared  to 
that of the original Hamiltonian. We apply this method in combination with the full configuration interaction quantum Monte Carlo (FCIQMC) method to the homogeneous electron gas. As a projection technique, the non-Hermitian nature of the transcorrelated Hamiltonian 
does not cause complications or numerical difficulties for FCIQMC. The rate of convergence of the total energy to the 
complete basis set limit is improved from ${\cal O}(M^{-1})$ to ${\cal O}\left({\MMF}\right)$, where $M$ is the total number 
of orbital basis functions.

\end{abstract}

\maketitle

\section{Introduction}  
\label{introduction}
Electron correlation can be roughly classified into static correlation and dynamic correlation. In conventional configuration descriptions of 
the  many-body wave function, the two different types of electron correlation are treated in the same way, i.e.,  by linear expansion  
in terms of Slater  determinants \cite{helgaker99}. While such a configuration description offers a natural and efficient way to deal with static correlation, 
it, however, does not treat dynamic  correlation efficiently, and thus usually leads to a slow convergence to the complete basis limit (CBL). 
The main problem is that, due to the Coulomb singularity of the electronic interaction, the short-range dynamic correlation introduces non-smoothness  
into the many-body wave function, which can not be approximated efficiently by orbital product expansions.
This problem is even more severe for those methods aiming at high accuracy, such as full configuration interaction (FCI) 
methods and high-order coupled-cluster methods.
On the other hand, the non-smoothness of the many-body wave function can be locally resolved and is expressed as the well known Kato cusp condition \cite{kato57}
\begin{equation}
  \left. \frac {\partial \Psi} {\partial r_{ij}}\right| _{r_{ij}=0}=\frac 1 2 \left. \Psi \right| _{r_{ij}=0}. 
   \label{eq_cusp}
\end{equation}
Incorporating this property into the construction of many-body wave functions should in principle speed up convergence to the CBL.

One way of incorporating the cusp condition is to introduce explicitly correlated basis functions for the expansion of the many body-wave 
function. In the last few decades,  various explicit correlation methods (e.g., R12  and F12 methods  \cite{kutzelnigg85, KK91, 
KMTV, HKKT, THBK, WAKM, Kong12}) have been suggested and have achieved a high level of success.  The main feature of these methods is that,  
instead of the conventional approximation of the cusp in terms of orbital product expansions, electron pair geminal functions are directly used in 
the construction of the basis  of  the many-body wave function. These geminal functions describe the cusps very efficiently, although they also make  the involved calculations highly non-linear. For example, in these calculations one has to deal with various kind of orthogonality 
constraints, which lead to multi-electron integrals (three, four and even higher-body electron integrals).  These integrals are usually 
approximated by  resolution of identity (RI) techniques \cite{kutzelnigg85, KK91, KMTV, valeev04}.  

Electron cusps can also be efficiently described by using the Jastrow ansatz \cite{jastrow55} 
\begin{equation}
\Psi({\bf R})=e^{\tau ({\bf R})} \Phi ({\bf R}), \qquad {\bf R}=({\bf r}_1,{\bf r}_2,\cdots,{\bf r}_N),
\label{eq_jast}
\end{equation}
where $\Phi$ is an anti-symmetric reference function and $\tau$ is a symmetric pair correlation factor
\begin{equation}
\tau ({\bf R})=\frac 1 2 \sum_{i \neq j}  u({\bf r}_i,{\bf r}_j).
\label{eq_tau}
\end{equation}
The correlation factor can be constructed to fulfill the cusp condition (\ref{eq_cusp}), and thus the regularity of the reference function $\Phi$ is higher than that of the wave 
function $\Psi$. Fournais {\it et al.} \cite{FHHO05} have proven that the correlation factor can improve the regularity of the wave function from $C^{0,1}$ to $C^{1,1}$. 
In Appendix \ref{app1} we will show that this will lead to a speedup of basis convergence for three dimensional non spin-polarised systems from $M^{-1}$ to $\MMF$. 
\footnote[5]{
It has been found that $C^{1,1}$ is the optimal regularity for the product ansatz, and thus we can only expect a $M^{-5/3}$ convergence for the Jastrow ansatz. 
However, it has been pointed out by  Fournais {\it et al.} that higher order regularity can be expected for a more general additive ansatz \cite{FHHO09}.
This is consistent with the early work of Kutzelnigg and Morgen \cite{kutzelnigg92}, where, based on such an additive ansatz, different cusp 
conditions are applied to different types of electron pairs, 
and thus a higher order convergence (such as $M^{-7/3}$) is achieved. Unfortunately such kind of additive ansatz is not size consistent.}
In order to guarantee size consistency, the correlation factor has to take an exponential form. 
This makes the Jastrow ansatz highly non-linear and any variational treatment 
leads to extremely high-dimensional integrals. Presently the Jastrow ansatz is primarily used in various quantum Monte Carlo methods, 
such as variational Monte Carlo (VMC) and diffusion Monte Carlo (DMC) methods \cite{ceperley78,umrigar93,foulkes01}, where the involved  
integrals can be evaluated directly in the high dimensional space. The exponential correlation factor has also been treated by various expansions, 
such as the linked cluster expansion \cite{Clark68, Talman74}, random phase approximations (RPA)  \cite{gaudoin01}, Fermi hypernetted chain  (FHNC) 
method \cite{krotscheck84,krotscheck85}. These sophisticated methods are highly nonlinear and difficult to implement in practical calculations.

For an efficient treatment of the exponential correlation factor, a relatively simple method, the transcorrelated (TC)  method \cite{Boys69, 
Boys69a, Boys69b, Boys69c,  Boys69d, Handy69, Handy72, Handy73}, was suggested by Boys and Handy roughly half a century ago.
By using a similarity transformation, $e^{-\tau}\hat{H}e^{\tau}$, the exponential correlation factor is removed from the involved equations.  
The original TC method of Boys and Handy was designed for the single-determinant Jastrow ansatz and contains two equations, for the calculation  
of the correlation factor $\tau$ and the orbitals respectively. Initial calculations demonstrated that 
this method can efficiently recover much of the correlation energy. On the other hand the resulting non-Hermitian effective Hamiltonian cannot prevent the 
energy from falling below the exact one. The lack of a variational bound is considered to a severe problem and has hampered  a broad 
application of the TC method for quite a long time. 

Recently, there has been renewed interest in the development of the TC 
method. Ten-no  used the TC Hamiltonian in the perturbation and the coupled electron-pair approximations \cite{Ten-No00}. In this 
approach, the correlation factor $\tau$  is a fixed local geminal satisfying the cusp condition, while the reference function is treated by 
conventional configuration expansions. The reference function $\Phi$ is much smoother than the many-body wave function
$\Psi$ and,  as a consequence,  the  configuration expansion of $\Phi$ converges much faster.  
The price to pay is the introduction of  a  three-body operator in the effective 
Hamiltonian, as well as various numerical problems due to non-Hermiticity. The non-Hermiticity problem is more  severe for self 
consistent  optimization methods. Hino   et al. suggested to use biorthogonal basis to deal with the TC Hamiltonian \cite{Hino01}.  
Umezawa et al.   optimized  the orbitals by minimizing the energy  variance \cite{Umezawa03,Umezawa05}. Luo  has suggested a general variational 
method for simultaneous  optimizations of  the correlation factor and the reference function \cite{Luo10a, Luo11}.  Yanai et al. have used a truncated 
canonically transformed Hamiltonian to eliminate the non-Hermiticity of the effective Hamiltonian\cite{Yanai06,Yanai12}.
Gr\"uneis et al. recently offered a detailed discussion on the choice of correlation factors \cite{Grueneis17}.

In this work we incorporate the TC method into the full configuration-interaction quantum Monte Carlo (FCIQMC)  method
\cite{BTA09,CBA2010,BGKA13}, aiming at  highly accurate calculations on periodic systems.  
FCI in principle provides the most accurate description of the wavefunction, within an orbital representation, 
and the results are usually used to benchmark  other calculation results.
However, FCI methods are also extremely expensive, with the computational cost scaling exponentially with respect to the system size.   On the 
other hand, the extremely large FCI expansion is also very sparse, so that the vast majority of expansion coefficients are essentially zero. This sparsity, 
however, usually has no regular pattern, especially for strongly correlated systems. The recently developed FCIQMC method and its `initiator' 
adaptation ($i-$FCIQMC)  \cite{CBA2010,BoothC2} offers a way to detect and make use of this sparsity. 
This method is based on Monte Carlo simulations of the dynamic evolution of the many-body wave function with imaginary time:
\begin{equation}
\Psi (t) = e^{-t (\hat{H}-E_0 )}\Psi(t=0),  
\label{eq_Psi_t}
\end{equation}
which leads to the ground state wave function in the long-time limit $\Psi_0=\Psi(t \rightarrow \infty)$.
The FCI expansion coefficients are simulated by a set of 
walkers which evolve over imaginary time. In the long-time limit, a steady distribution of the walkers is reached and the corresponding
projection energy, in the large-walker limit, converges to the FCI energy. With this method, FCI-quality calculations have been achieved on larger molecular systems \cite{KBA2016} and even 
periodic systems \cite{BGKA13}.
F12 methods have also been combined with the FCIQMC technique \cite{BCAT2012,KBA2016} as universal {\em a posteriori} 
corrections \cite{TorheydenValeev09, KongValeev11} involving contractions of the 
the one- and two-body matrices with F12 integrals.  

For periodic systems, plane waves are usually the most appropriate basis functions. However their smooth and non-local properties make the slow
 convergence in the description of electronic cusps an even more severe problem. This problem is tremendously enhanced in FCIQMC with large basis sets, aiming to reach the CBL. 
 In this work we design an explicit correlation method to resolve this problem in such calculations.  In FCIQMC, the matrix 
 elements of the interaction operators (e.g., the two-body Coulomb operators, etc.) are used intensively and thus have to be either stored 
 efficiently or calculated repeatedly on the fly.   One advantage of the plane wave basis is that the two-body Coulomb matrix 
 can be simply evaluated  and need not be stored, alleviating the memory bottlenecks associated with storing the 4-index integrals
 of large systems.  We would like to keep this advantage, so 
 that the application of  the intended explicit correlation method will not be limited to small systems.  This requires that the involved effective 
 potential and their matrix elements should be as simple as possible. In the next section, we will describe the new TC
 method  designed for plane wave basis. In section \ref{results}, we will present our initial test calculations on three dimensional homogeneous electron 
 gas models and, this will be followed by some conclusions and discussions in section \ref{conclusion}. 

\section{Method}
\label{method}
Following the idea of Ten-no \cite{Ten-No00}, for a given $N$ electron system with a given basis set,  we take a fixed correlation factor $\tau$ in the Jastrow 
ansatz (\ref{eq_jast}) and,  try to determine the reference function $\Phi$ by solving approximately  an eigenvalue equation
\begin{equation}
\hat{H}_{\TC}\Phi =E \Phi,
\label{eq_eigenvalue}
\end{equation}
where the effective Hamiltonian, the transcorrelated Hamiltonian, has a finite Baker-Campbell-Hausdorff expansion, upto a double commutator, owing 
to the fact that the correlation factor $\tau({\bf R})$ is purely a function of the spatial coordinates of the electrons: 
\begin{eqnarray}
\hat{H}_{\TC}&\equiv& e^{-\tau}\hat{H}e^{\tau} \nonumber \\
&  =& \hat{H}+[\hat{H},\tau]+\frac 1 2 [[\hat{H},\tau],\tau] \nonumber \\
&=& \hat{H}- \sum_i \left( \frac 1 2 \triangledown_i ^2 \tau +(\triangledown_i \tau)\cdot \triangledown_i +\frac 1 2 (\triangledown_i \tau)^2  \right).
\label{eq_HTC}
\end{eqnarray}
Here the TC Hamiltonian is non-Hermitian, owing to the presence of the single commutator term $[\hat{H},\tau]$.   For any 
eigenvalue of such operators,  the corresponding left and right eigenvectors are usually different.  In solving this type of eigenvalue equations,
 approximations based on a  variational treatment  are  usually troubled by the non-Hermitian nature of the matrix.  Projection methods, however,
such as the Power method provide a route forward for such matrices \cite{Golub}. 
Unlike standard FCI methods, the FCIQMC method is not a variational method but rather a stochastic version of the Power method. 
 For the  Jastrow ansatz,  the time evolution of the wave function can  be represented  as
\begin{eqnarray}
\Psi(t)&=& e^{\tau}\Phi(t), \\
\Phi(t)&=&e^{-t (\hat{H}_{\TC}-E_0 )}\Phi(t=0),  \label{eq_Phi_t}
\end{eqnarray}
where equation (\ref{eq_Phi_t}) can be simply derived from equation (\ref{eq_Psi_t})
\begin{eqnarray}
\Phi(t)&=& e^{-\tau}e^{-t (\hat{H}-E_0 )}e^{\tau} \Phi(t=0)  \nonumber \\
           &=&  \lim_{M\rightarrow \infty}e^{-\tau} \left(1-\frac  {t (\hat{H}-E_0 )} M \right) ^M e^{\tau} \Phi(t=0), \nonumber \\
           &=&  \lim_{M\rightarrow \infty}\left(e^{-\tau} \left(1-\frac  {t (\hat{H}-E_0 )} M \right)  e^{\tau}\right)^M \Phi(t=0), \nonumber \\
           &=&  \lim_{M\rightarrow \infty}\left(1-\frac  {t (\hat{H}_{TC}-E_0 )} M \right) ^M  \Phi(t=0), \nonumber \\
           &=& e^{-t (\hat{H}_{\TC}-E_0 )}\Phi(t=0).
\end{eqnarray}
It is worth noticing that equation (\ref{eq_Phi_t}) is not constructed based on the eigenvalue equation (\ref{eq_eigenvalue}), where one may get 
frustrated due to the non-Hermiticity  and lack of  variational bounds on $\hat{H}_{\TC}$.  The equivalence of  equation (\ref{eq_Psi_t}) and equation 
(\ref{eq_Phi_t}) reveals  that methods based on these equations (such as  FCIQMC)  can handle properly the non-Hermiticity due to such kind of similarity 
transformations.  Following equation (\ref{eq_Phi_t}), the FCIQMC method can be directly used for the TC Hamiltonian $\hat{H}_{\TC}$. 
The only difference is that here we are dealing with non-Hermitian operators, so that the involved matrix elements are non-symmetric. In calculations of 
these matrix elements,   the operators should not be mixed up with their Hermitian conjugates (i.e., these operators should only be applied 
on to the right hand side).

For periodic systems, the orbital basis functions to be used are plane waves,  
\begin{equation}
\phi_{{\bf p},\sigma}({\bf r})= \frac 1 {\sqrt{\Omega_0}}e^{i{\bf p}\cdot { \bf r}},
\label{eq_planewave}
\end{equation}
where $\Omega_0 $  is the formal volume of the infinite system.  All operators in the effective Hamiltonian can then be 
represented in terms of second quantisation  \cite{Luo12}.  The fixed correlation factor, assumed to be spin-independent, can 
be expressed as
\begin{eqnarray}
\tau &=& \frac 1 2 \sum_{ij} 
u_0({\bf r}_i-{\bf r}_j) \nonumber\\
&=& \frac 1 2 \sum_{\sigma \sigma'} \sum _{{\bf k}{\bf p}{\bf q}} 
\langle {\bf p}-{\bf k},{\bf q}+{\bf k}|u_{0}|{\bf p},{\bf q} \rangle  
a^{\dagger}_{{\bf p}-{\bf k},\sigma} a^{\dagger}_{{\bf q}+{\bf k},\sigma'}a_{{\bf q},\sigma'}a_{{\bf p},\sigma}\nonumber\\
&=&\frac 1 {2\Omega_0} \sum_{\sigma \sigma'} \sum _{{\bf k}{\bf p}{\bf q}} {\tilde u}_{0}({\bf k})
a^{\dagger}_{{\bf p}-{\bf k},\sigma} a^{\dagger}_{{\bf q}+{\bf k},\sigma'}a_{{\bf q},\sigma'}a_{{\bf p},\sigma},
\label{eq_tau2}
\end{eqnarray}
where ${\tilde u}_{0}({\bf k})=\int e^{i{\bf k}\cdot {\bf r}}u_0({\bf r})d^3 r$ is the Fourier transformation of $u_0$.  Here the two-body correlation factor
is assumed to be a function of ${\bf r}_{ij}={\bf r}_i-{\bf r}_j$, due to translational symmetry.  Similarly, for the other 
required two-body operators, 
we have the following expressions
\begin{eqnarray}
\hat{W} & = & \frac{1}{2\Omega_{0}}\sum_{\sigma\sigma'}\sum_{\bf k p q}\ \tilde{w}_{0}({\bf k})\ a_{{\bf p}-{\bf k},\sigma}^{\dagger}a_{{\bf q}+{\bf k},
\sigma'}^{\dagger}a_{{\bf q},\sigma'}a_{{\bf p},\sigma},\\
\frac{1}{2}\sum_{i}\triangledown_{i}^{2}\tau&=& -\frac{1}{2\Omega_0}\sum_{\sigma\sigma'}\sum_{\bf k p q}k^{2}\tilde{u}_0 ({\bf k})a_{{\bf p}-{\bf k},
\sigma}^{\dagger}a_{{\bf q}+{\bf k},\sigma'}^{\dagger}a_{{\bf q},\sigma'}a_{{\bf p},\sigma},\\
\sum_{i}(\triangledown_{i}\tau)\triangledown_{i} & = &\frac{1}{2\Omega_0}\sum_{\sigma\sigma'}\sum_{\bf k p q}({\bf p}-{\bf q})\cdot {\bf k}\ 
 \tilde{u}_{0}({\bf k})a_{{\bf p}-{\bf k},\sigma}^{\dagger}a_{{\bf q}+{\bf k},\sigma'}^{\dagger}a_{{\bf q},\sigma'}a_{{\bf p},\sigma},
\end{eqnarray}
where $\hat {W}$ is the electronic (Coulomb) potential.
The double commutator in equation (\ref{eq_HTC}) is more complicated and gives rise to a three-body operator and a two-body operator
\begin{eqnarray}
\frac{1}{2}\sum_{i}(\triangledown_{i}\tau)^{2} &=&
\frac{1}{2}\sum_{ijk}\triangledown_{i}u_0 ({\bf r}_{i}-{\bf r}_{j}) \cdot \triangledown_{i }u_0 ({\bf r}_{i}-{\bf r}_{k}) 
+\frac{1}{2}\sum_{ij} \left( \triangledown_i u_0 ({\bf r}_{i}-{\bf r}_{j}) \right)^2  \nonumber \\
&=& 
\frac{1}{2\Omega_0^{2}}\sum_{\sigma\sigma'\sigma''}\sum_{\bf k k' p q s} \tilde{u}_0({\bf k}) \tilde{u}_0({\bf k}'){\bf k}'\cdot {\bf k}\ a_{{\bf p}-{\bf k},
\sigma}^{\dagger}a_{{\bf q}+{\bf k}',\sigma'}^{\dagger}a_{{\bf s}+{\bf k}-{\bf k}',\sigma''}^{\dagger}a_{{\bf s},\sigma''}a_{{\bf q},\sigma'}a_{{\bf p},
 \sigma} \nonumber \\
&&+\frac{1}{2\Omega_0}\sum_{\sigma\sigma'}\sum_{\bf k p q}  \mathcal{F}((\triangledown u_0)^{2}) ({\bf k})\ a_{{\bf p}-{\bf k},
\sigma}^{\dagger}a_{{\bf q}+{\bf k},\sigma'}^{\dagger}a_{{\bf q},\sigma'}a_{{\bf p},\sigma},
\label{eq_DC}
\end{eqnarray}
where $\mathcal{F}$ denotes Fourier transformation.  A complete treatment of the three-body operator would be 
expensive, and we would like to treat it only approximately.

Periodic systems  are usually  treated by the supercell approach \cite{foulkes01},  where the infinite system is approximated by  periodically 
arranged replicas of a  finite cell $\Omega=L^3$.  Due to this artificial periodic boundary condition, the ${\bf p}$ vector of the plane wave basis 
(\ref{eq_planewave}) is discretized (i.e., ${\bf p}=\frac {2\pi} L {\bf n}, \ {\bf n}\in  {\cal Z}^3$), and to make the basis finite we take a cutoff
 $|{\bf p}|\le k_c$.  

In the supercell approach, $u_0$ and $w_0$  have to fulfill the periodic boundary conditions. The periodic $w_0$ is usually 
constructed via periodic summation\cite{foulkes01}.
Following the same idea,  we construct the periodic correlation factor $u_{0}$ by applying the periodic summation on a local function $u$
\begin{equation}
u_{0}({\bf r})=\sum_{{\bf {\bf n}\in{\cal Z}^3}}u({\bf r}+{\bf n}L).
\end{equation}
However such summations are not practically needed if we work in k-space, since
\begin{equation}
\frac{\tilde{u}_{0}({\bf k})}{\Omega_{0}}=\frac{\tilde{u}({\bf k})}{\Omega},\qquad \mbox{for } {\bf k}=\frac{2\pi}{L}{\bf m},\ {\bf m}\in\mathcal{Z}^{3}.
\end{equation}
The only thing which needs to be taken care of is that the inverse Fourier transformation of $u_{0}$ does not exist, and instead,  there is a Fourier series
\begin{equation}
u_{0}({\bf r}_{2}-{\bf r}_{1})=\frac{1}{\Omega}\sum_{{\bf k}=2\pi{\bf m}/L}\tilde{u}({\bf k})e^{i{\bf k}\cdot({\bf r}_{2}-{\bf r}_{1})}.
\end{equation}
The two-body operator $(\triangledown u_0)^2$ is very important for the short range correlation and, it is closely related to the infinite summation of all
ladder diagrams in the linked cluster expansion \cite{Talman74}.  By taking the periodic summation,  $(\triangledown u_0)^2$ can be expressed as
\begin{equation}
(\triangledown u_{0})^{2} =-\frac{1}{\Omega^{2}}\sum_{{\bf k},{\bf k}'}{\bf k}\cdot{\bf k}'\tilde{u}({\bf k})\tilde{u}({\bf k}')e^{i{\bf k}\cdot({\bf r}_{2}-{\bf 
r}_{1})}e^{i{\bf k}'\cdot({\bf r}_{2}-{\bf r}_{1})},
\end{equation}
whose Fourier transformation is now 
\begin{equation}
{\cal \mathcal{F}}((\triangledown u_{0})^{2})({\bf k})=-\frac{\Omega_{0}}{\Omega}\left(\frac{1}{\Omega}\sum_{{\bf k}'}({\bf k}-{\bf k}')\cdot{\bf k}'\tilde{u}
({\bf k}-{\bf k}')\tilde{u}({\bf k}')\right).
\end{equation}

\begin{figure}[h]  
\centering        
\begin{overpic}[scale=1,tics=10]
{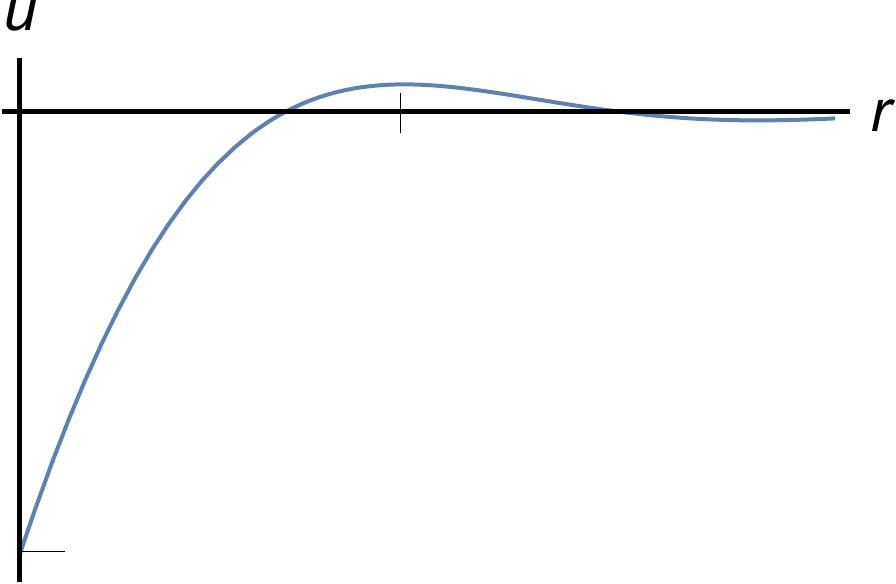}
\put(43,43){\Large $\frac \pi  k_c$}
\put(10,2){\Large $-\frac 2 {\pi k_c}$}
\end{overpic}
\caption{ A sketch of the correlation factor  $u(r)$. }
\label{fig_ur}    
\end{figure}

As explained before, the correlation factor $\tau$ is designed here mainly to capture the short range cusp. 
In the short range limit, we have the asymptotic solution \cite{Luo12}
\begin{equation}
\tilde{u}({\bf k})=-\frac{4\pi}{{\bf k}^{4}},\ \mbox{when} \ k\sim\infty,
\end{equation}
which is  the cusp condition expressed in the  k-space.
This expression is derived for unlike spin pairs, while for electron pairs with the same spin $\tilde{u}({\bf k})\sim-\frac{2\pi}{{\bf k}^{4}}$.
In principle we can use any kind of local function $u$ in the TC calculation, as long as it  satisfies the
cusp condition. However, since we intend to introduce approximations in the treatment of the three-body term, we want $u$ to be small
and to vanish in the CBL. Therefore we design the following correlation factor
\begin{equation}
\tilde{u}({\bf k})=\begin{cases}
-\frac{4\pi}{k^{4}}, & |{\bf k}|>k_{c},\\
0, & |{\bf k}|\le k_{c},
\end{cases}
\label{eq_uk}
\end{equation}
where $k_{c}$ is the cutoff parameter of the basis set.  The idea behind this construction is simple:  since the wave function can already be
described by a configuration description up to the given level of resolution (characterized by $k_c$),  it is only needed to be improved in the finer region of resolution by means of
the correlation factor.  In real space the above correlation factor becomes: 
\begin{equation}
u(r)=-\frac{r}{\pi}\left(\mbox{si}(k_{c}r)+\frac{\cos(k_{c}r)}{k_{c}r}+\frac{\sin(k_{c}r)}{(k_{c}r)^{2}}\right),
\label{eq_ur}
\end{equation}
where $\mbox{si}(x)=-\int_x ^\infty \frac{\sin x}{x} dx$ is the sine integral.  A sketch of $u(r)$ is presented in Figure \ref{fig_ur},  which looks like a small ``hole" 
with the $\mbox{depth}=-2/ {\pi k_c}$ and the $\mbox{width}\sim  \pi/ k_c$.
Taylor expansion of $u$ in the small $r$ region can be calculated for the leading terms
\begin{equation}
u(r)=-\frac{2}{\pi k_{c}}+\frac{r}{2}+\cdots,
\end{equation}
where  the second term satisfy the cusp condition for unlike spin pairs. In the large $r$ region, the magnitude of $u(r)$ decades like $1/r^2$. 

In principle, we can also use a spin-dependent correlation factor, namely to reduce the correlation factor for parallel spin pairs by one half. However,
the existence of exchange ``holes'' between parallel-spin electrons keeps them largely apart anyway, and as a consequence
the system energy is not very 
sensitive to the cusp between such pairs.  Therefore the use of spin-dependent correlation factors does not significantly 
improve the convergence rate of energy. Rather, 
spin-dependent Jastrow factors induce undesirable spin contamination \cite{Filippi96} into the wave function, so that the wave function can not be an eigenstate of $S^2$.

With the above short ranged correlation factor, we can largely ignore the complicated three-body operator in equation (\ref{eq_DC}) and take
only a simple RPA type contribution from it. This contribution is represented by a two-body operator 
\begin{equation}
\frac{N-2}{2\Omega^2}\sum_{\sigma\sigma'}\sum_{\bf k  p q } k^2 \tilde{u}^2({\bf k}) \ a_{{\bf p}-{\bf k},
\sigma}^{\dagger}a_{{\bf q}+{\bf k},\sigma'}^{\dagger} a_{{\bf q},\sigma'}a_{{\bf p},
 \sigma},
\end{equation}
which is generated by a contraction of the $a_{{\bf s}+{\bf k}-{\bf k}',\sigma''}^{\dagger}a_{{\bf s},\sigma''}$ pairs in equation (\ref{eq_DC})
(i.e., by a summation of  those terms where ${\bf k}={\bf k}'$). This term makes the dominant contribution of the three-body operator to the long range correlation, and is 
closely related to the summation of all ring diagrams in the linked cluster expansion \cite{Armour80,Gaskell61,Gaskell62}. 
In the current method, the correlation factor is short ranged and, therefore the contribution of these terms will be very small. In the applications studied
in this paper, namely the homogeneous electron gas in the $r_s$ range from 0.5 to 5, this contribution was less than 1\% of the total correlation energy. 
We nevertheless keep it in the method  
for two reasons: firstly it partly recovers the three-body contributions and, secondly it can be used to estimate the magnitude of 
error due to the missing three-body terms. Last but not least, the use of this term incurs almost no extra computational cost.      

Putting all terms together, we have the following two-body effective potential
\begin{eqnarray}
\hat{W}_{\eff} & = & \frac{1}{2\Omega}\sum_{\sigma\sigma'}\sum_{\bf k p q}\ \tilde{w}_{\eff}({\bf k, p, q})\ a_{{\bf p}-{\bf k},\sigma}^{\dagger}a_{{\bf q}+{\bf k},
\sigma'}^{\dagger}a_{{\bf q},\sigma'}a_{{\bf p},\sigma},\\
\tilde{w}_{\eff}({\bf k, p, q})&=&\tilde{w}({\bf  k})+k^{2}\tilde{u}(k)-({\bf p}-{\bf q})\cdot {\bf k}\tilde{u}({\bf k})  
-\frac{N_{e}-2}{\Omega}k^{2}\tilde{u}^{2}({\bf k})  
\nonumber \\
 &  & +\frac{1}{\Omega}\sum_{{\bf k}'}({\bf k}-{\bf k}')\cdot{\bf k}'\tilde{u}({\bf k}-{\bf k}')\tilde{u}({\bf k}'), \label{eq_weff}
\end{eqnarray}
where $\tilde{w}({\bf k})$ is the original Coulomb potential
\begin{equation}
\tilde{w}({\bf k})=\begin{cases}
\frac{4\pi}{k^2}, & {\bf k}\neq {\bf 0},\\
0, & {\bf k}= {\bf 0}.
\end{cases}
\end{equation}
In equation (\ref{eq_weff}), the last term coming from   $\mathcal{F}((\triangledown u)^{2}) ({\bf k})$ is only a 
function of ${\bf k}$ and can be easily prepared and stored before the FCIQMC simulation.   
Unlike other terms, this term is non-zero at ${\bf k}={\bf 0}$.    Practical implementation of this TC method 
in FCIQMC calculations is straightforward. We need only to replace the pure electronic Coulomb
potential $\tilde{w}$ with the TC effective potential $\tilde{w}_{\eff}$,  and this requires only a very small modification of the existing 
code. Since calculation of the Coulomb potential  makes up only a very small portion of the total computational cost  of the  FCIQMC method, 
use  of the effective potential will not make the computations more expensive.

\section{Results}
\label{results}
Homogeneous electron gases (HEG) are important models for the investigation of electron correlation in solids. They also play a fundamental
role in the development of density functional theory (DFT) \cite{kohn65}. These models have been intensively  studied by variational and diffusion quantum
 Monte Carlo simulations \cite{Ceperley80, fraser96, Kwon98, Rios06, Drummond08}. 
 The DMC method is very efficient and can be applied to fairly large systems. On the other hand
  this method does not solve the problem completely, owing to the fixed-node approximation, 
even though this can be 
reduced by means of back-flow and multi-determinant techniques. The FCIQMC method contains in principle no bias, and offers a way to investigate the fixed node error.  
Recently the FCIQMC method has been used to investigate the three dimensional HEG model \cite{SBGA12, SBA12,NeufeldThom2017}.  Due to the cusp singularity of the wave 
functions, the calculation suffers from slow convergence with respect to the basis size, with the error  of the calculated energy being proportional to $\MMO$, $M$ being the total number of basis functions.  Extrapolations based on this $\MMO$ behavior are used to estimate the results in the CBL. For an
 accurate result, calculations need to be performed on fairly large basis sets in order to reach the $M^{-1}$ regime, and the computational cost increases sharply
 with basis size.

In order to study the efficiency of the new TC method, we have performed calculations on the same 3D HEG systems as investigated in
 the previous  studies \cite{SBGA12, SBA12}. Two different supercell sizes are used, containing   $14$ and $54$ electrons respectively. 
 For the $14$ electron cell, calculations are carried out on  four different densities with Wigner-Seitz radius $r_s=0.5$, $1.0$, $2.0$ and $5.0$ respectively. 
 For the $54$ electron cell, we can only get converged results for $r_s=0.5$ and $1.0$,  since the
 required  total number of walkers increases rapidly with $r_s$.   We use the initiator-FCIQMC ($i$-FCIQMC) method with the initiator   parameter set to be $3.0$. In addition we used the semi-stochastic method \cite{Petruziolo2012} using the $|D|=10000$ leading determinants in the deterministic space \cite{Blunt2015} as implemented in the NECI code \cite{BoothSmartAlavi2014}.

\begin{figure}[h]
\centerline{
\includegraphics[width=0.5\textwidth]{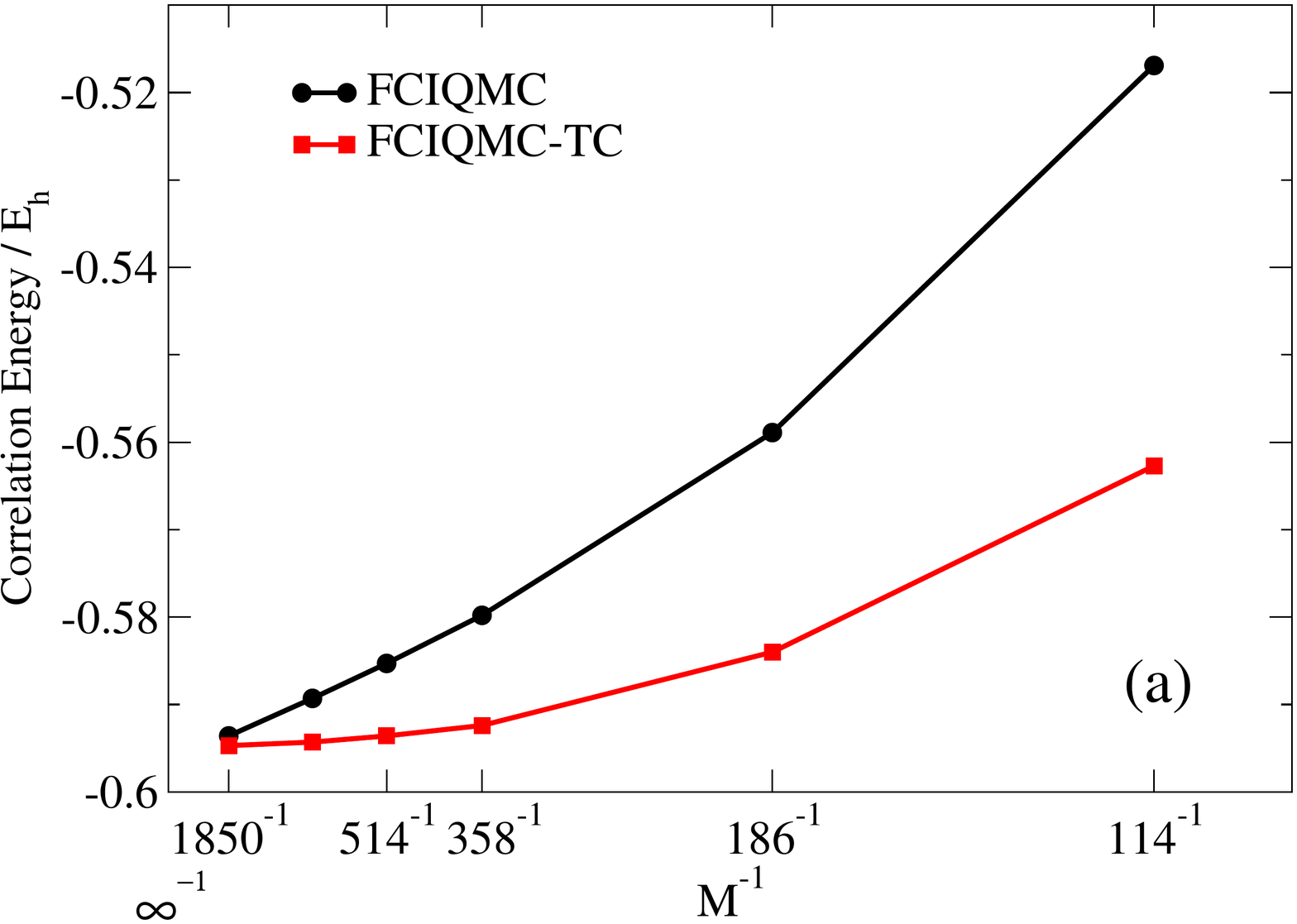}
\includegraphics[width=0.5\textwidth]{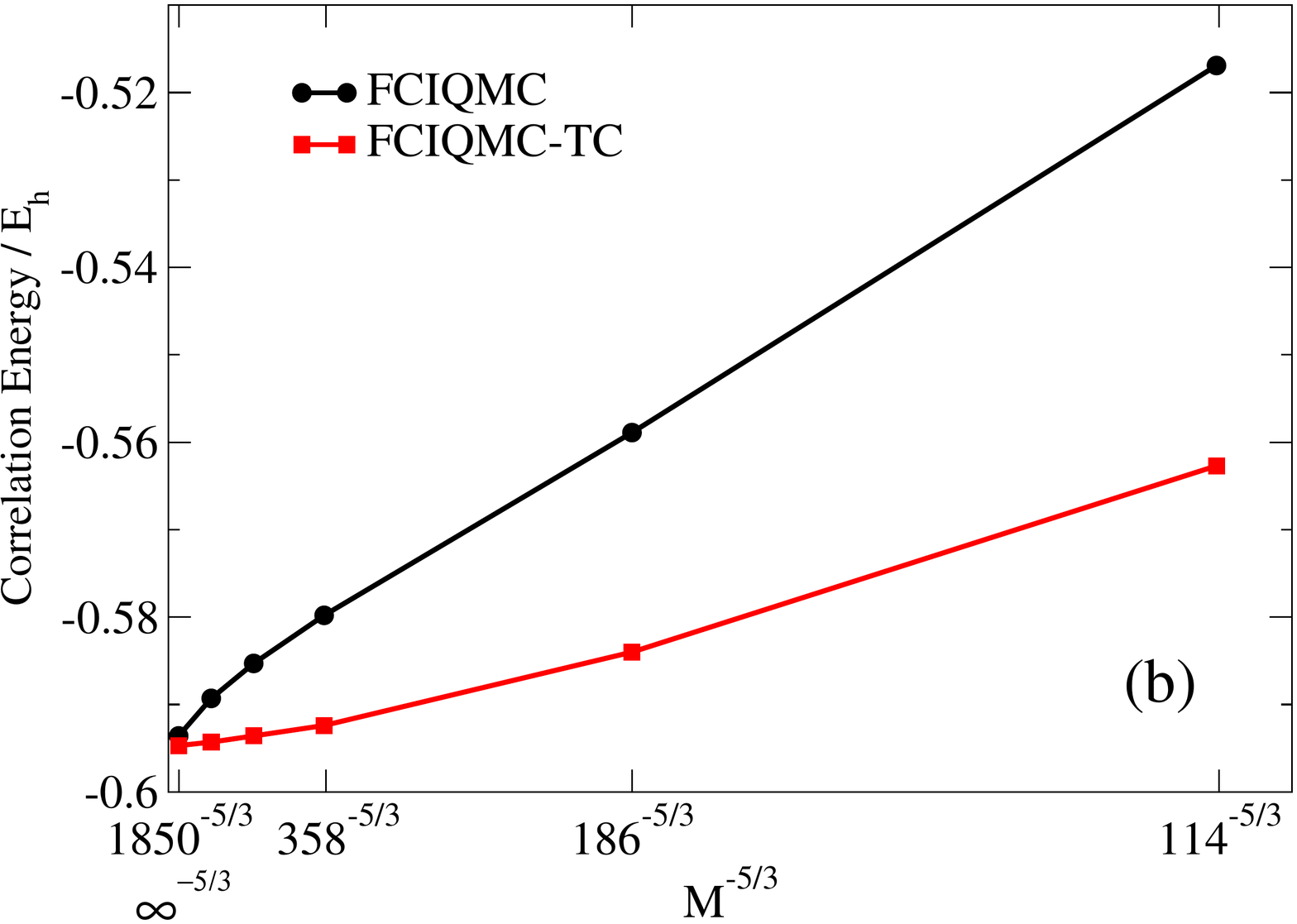}
}
\caption{(a) Total correlation energy as a function of $M^{-1}$ calculated by FCIQMC and  FCIQMC-TC methods for the 14 electron system with $r_s=0.5$. 
  (b) The same results  but presented as a function of $\MMF$.  }
\label{fig_0.5N14}
\end{figure}

 In Figure \ref{fig_0.5N14}(a),  the total correlation energy as a function of $M^{-1}$ is presented for the $14$ electron system with $r_s=0.5$, where the two different
 results are calculated by FCIQMC method with the effective TC Hamiltonian  (FCIQMC-TC) and   the original FCIQMC method respectively.  
Here  $M$,  the number of spin orbital basis functions,  is chosen to be $114$,  $186$, $358$, $514$, $778$ and $1850$ respectively.
The FCIQMC result shows an  asymptotic linear convergence with respect to $M^{-1}$, 
while the result of FCIQMC-TC has a higher order of convergence. According to the theoretical analysis in Appendix \ref{app1}, the best asymptotic convergence we can expect is
$\MMF$. We also present the same results in Figure \ref{fig_0.5N14}(b) as functions of $\MMF$. The result of FCIQMC-TC shows a roughly linear behavior in 
the large $M$ region, and in the small $M$ region the convergence is faster. 
 The asymptotic $\MMF$ convergence behavior offers a possibility of extrapolations to the CBL, in case of need. 
 For the result in Figure  \ref{fig_0.5N14}, such an extrapolation is not necessary, since $1850^{-5/3}$ is already very close to the origin ($\infty^{-5/3}$) and 
the FCIQMC-TC result has already converged at $0.1$ milihartree ($mE_h$) level.

\begin{figure}[h]
\centerline{
\includegraphics[width=0.333\textwidth]{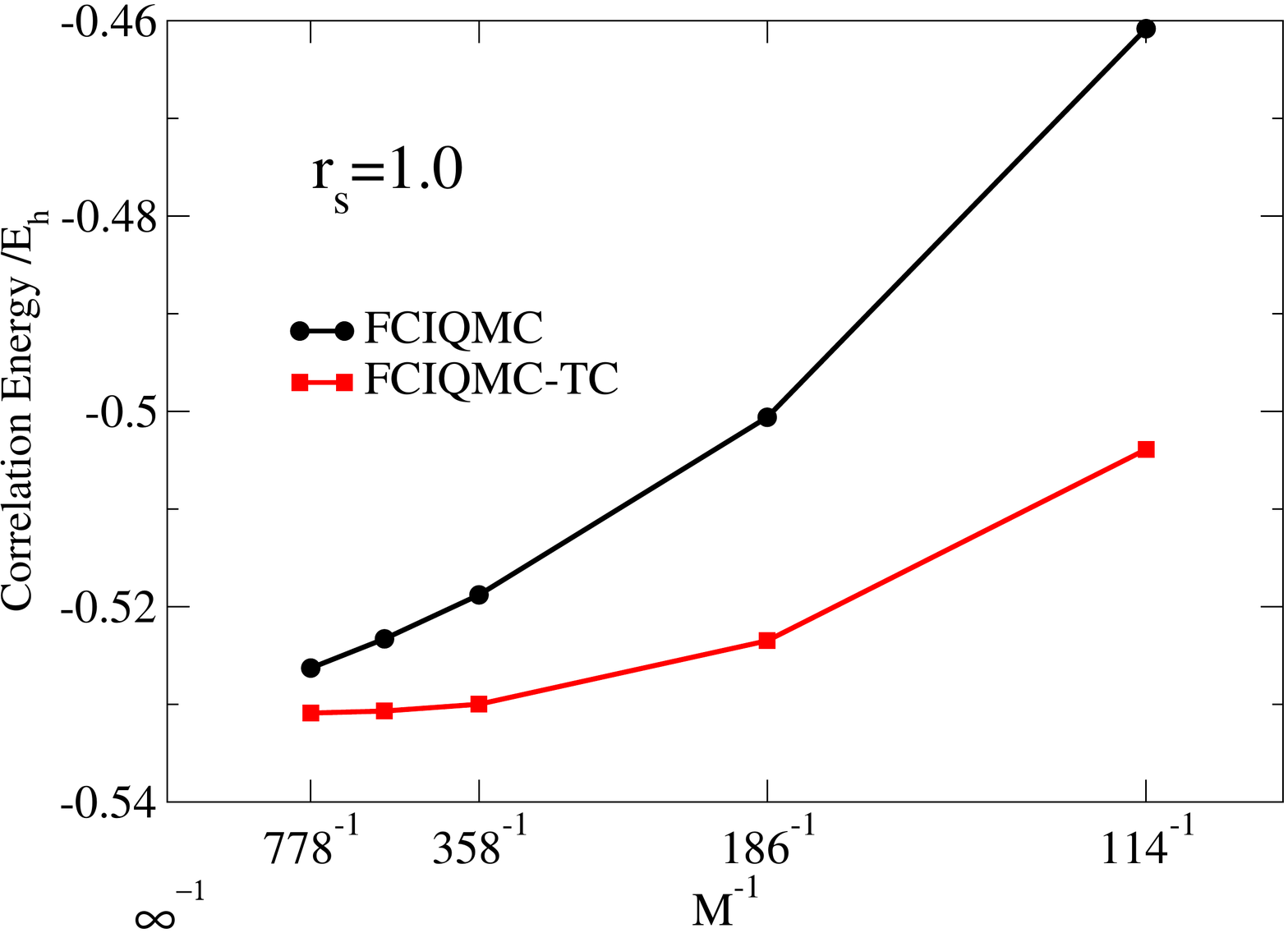}
\includegraphics[width=0.333\textwidth]{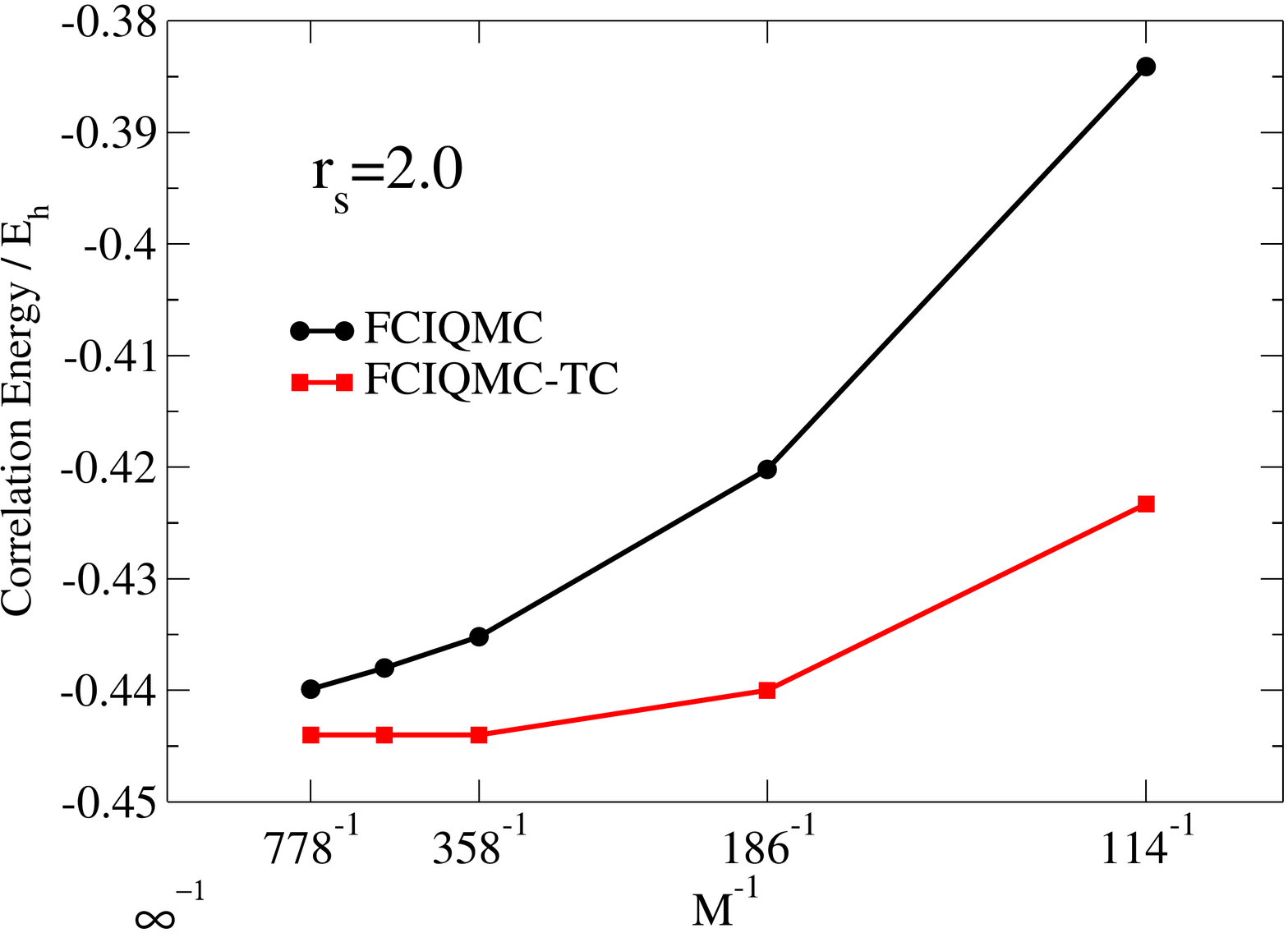}
\includegraphics[width=0.333\textwidth]{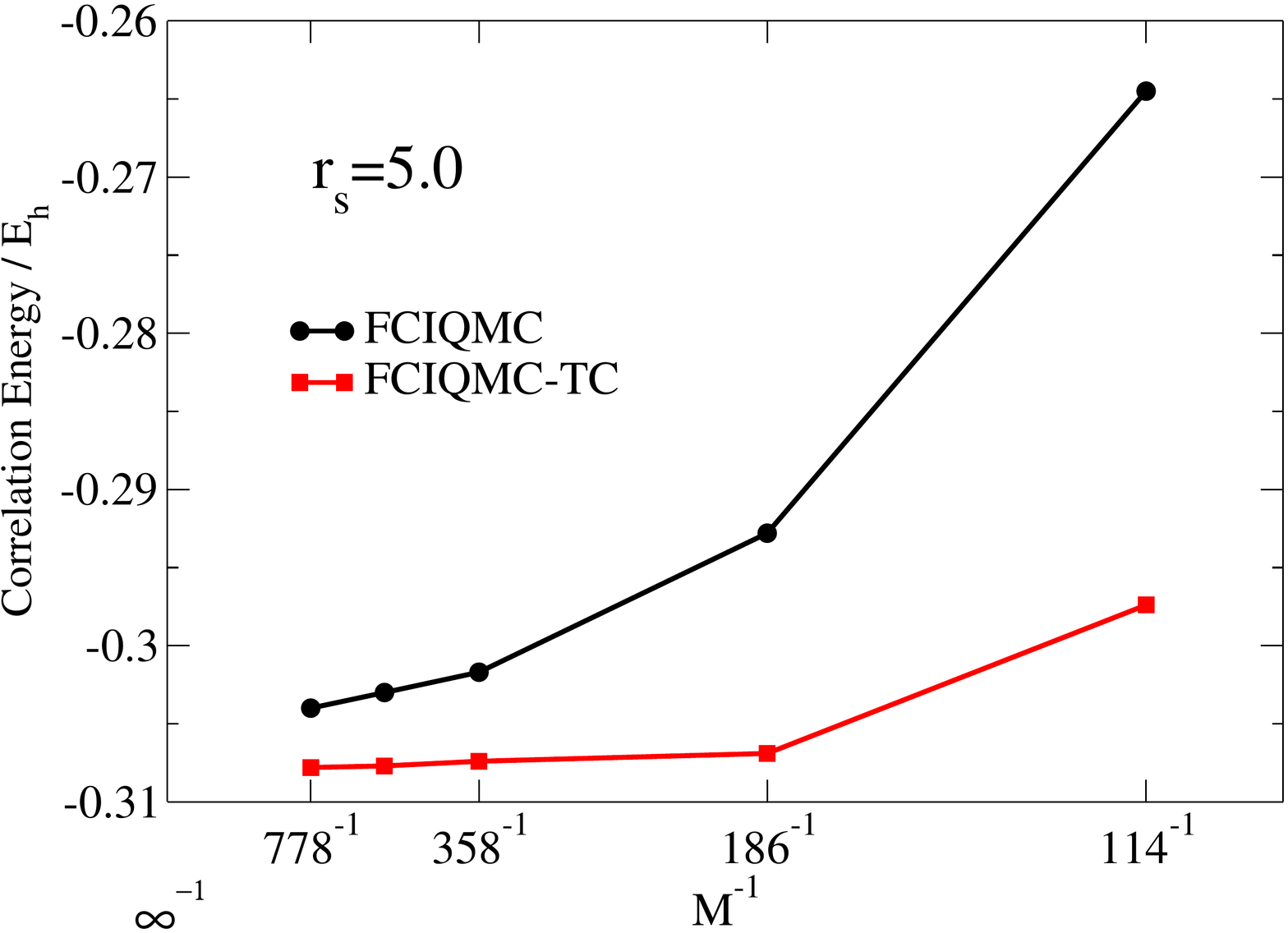}
}
\caption{Total correlation energy as a function   of $M^{-1}$ calculated by FCIQMC and  FCIQMC-TC methods for the 14 electron system with $r_s=1.0$,
$2.0$ and $5.0$ respectively.   }
\label{fig_N14}
\end{figure}

Similar behaviors are observed  for other densities of the 14 electron system.  In Figure~\ref{fig_N14}, the convergence with respect to $M^{-1}$ of the 
two different calculations are presented for $r_s=1.0$, $2.0$ and $5.0$ respectively in three plots.  It turns out  that for larger $r_s$ the result converges
faster.  For $r_s=1.0$ the FCIQMC-TC energy has converged within a milihartree error at $M=514$, while for $r_s=2.0$ and $5.0$ such a convergence can 
already been reached at $M=358$ and $186$ respectively.  However, this does not mean that simulations for larger $r_s$s are easier, since the required 
number of walkers ($N_w$) increases sharply with  $r_s$, in order to reduce the initiator error. For  $r_s=0.5 \sim 2.0$, calculations
are performed mostly with $N_w=10^7 \sim 10^8$, while for $r_s=5.0$ we need  $N_w=10^9$ already at  $M=114\sim 358$ and even $N_w=10^{10}$ at
$M=514$ and $778$.  For such systems, in order to demonstrate the $\MMF$ asymptotic convergence we have to deal with larger basis sets, which will require even larger
$N_w$'s, but on the other hand such difficult calculations are not needed for our accuracy requirement. 

\begin{figure}[h]
\centerline{
\includegraphics[width=0.5\textwidth]{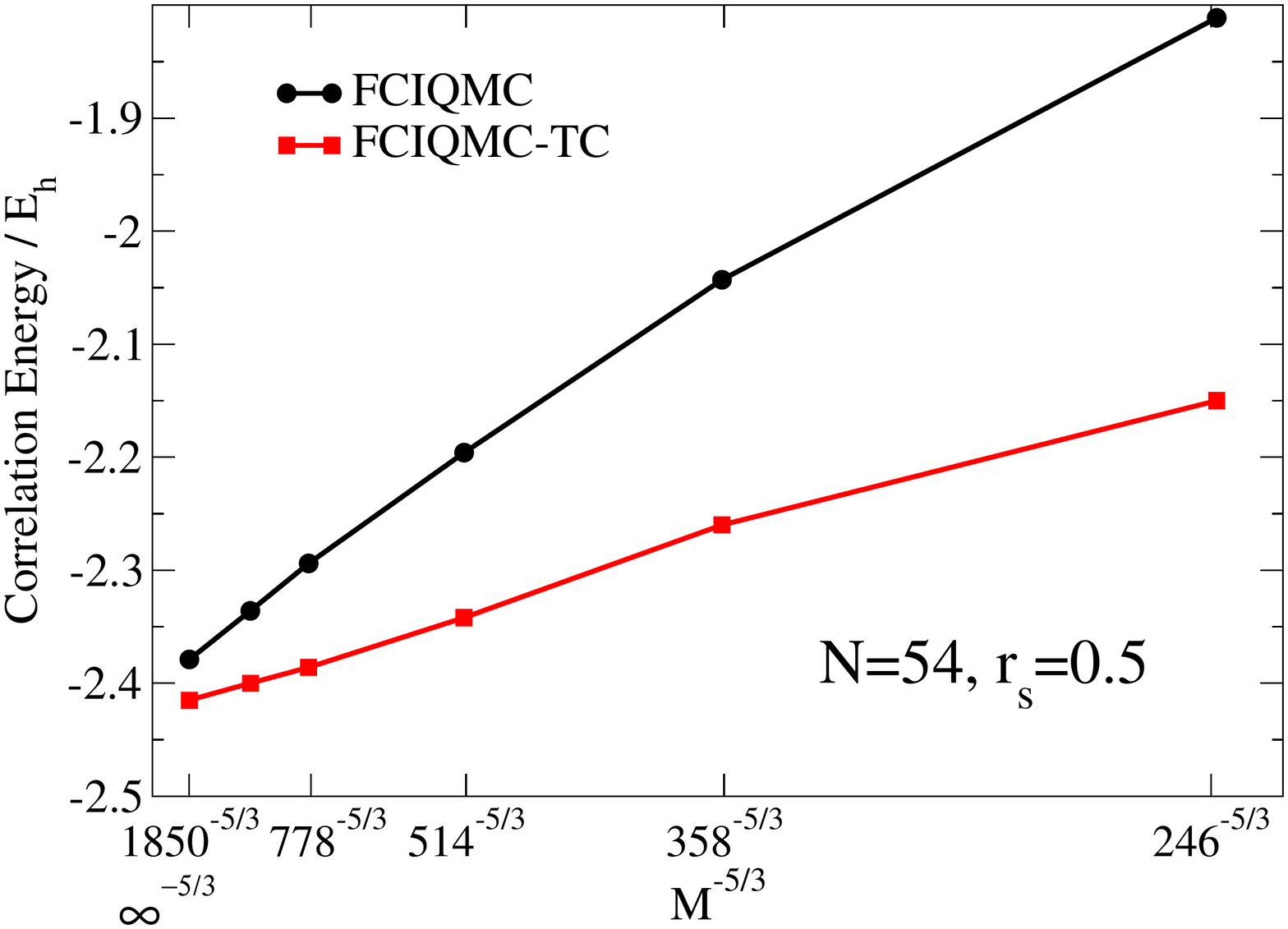}
\includegraphics[width=0.5\textwidth]{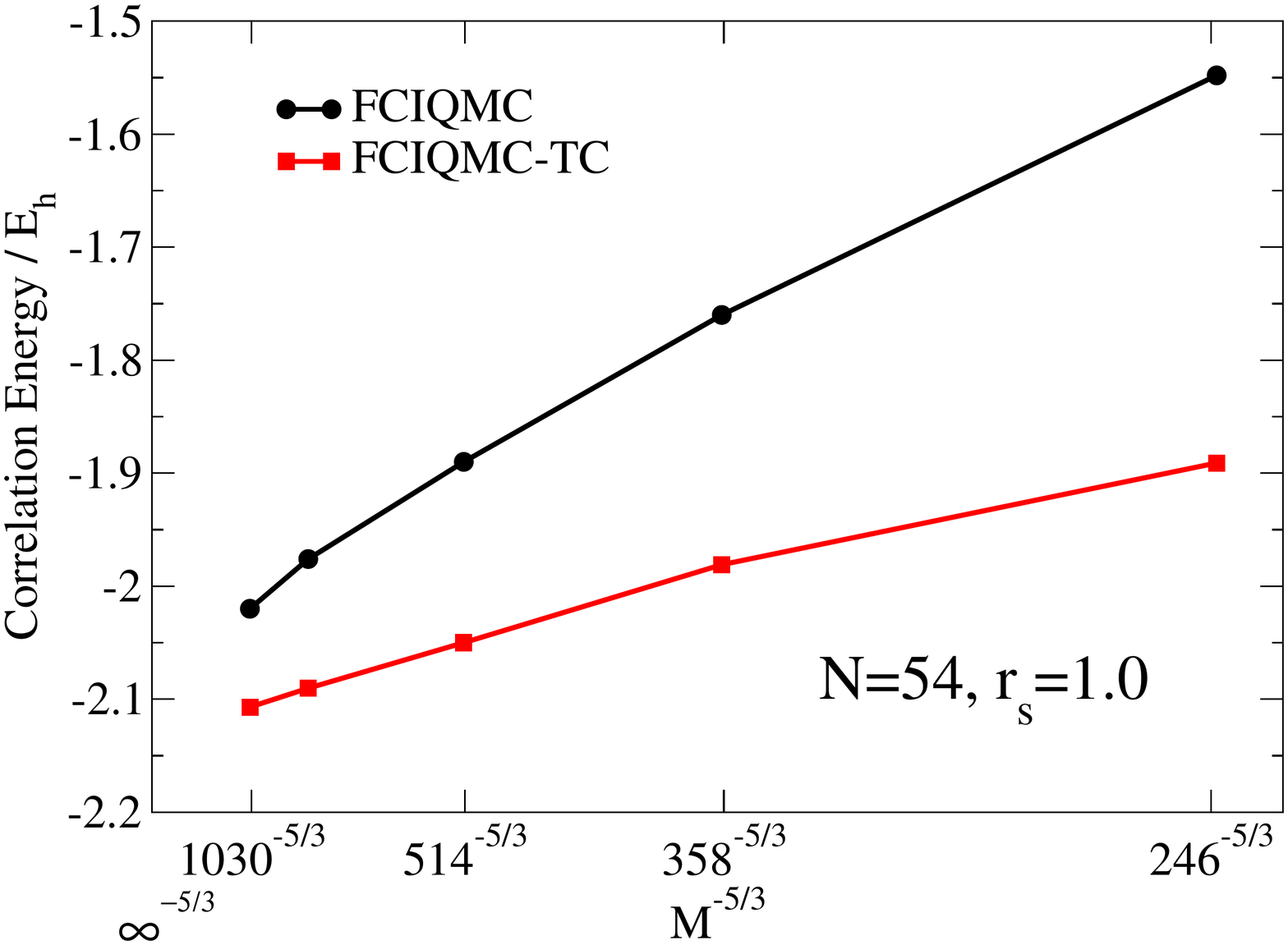}
}
\caption{Total correlation energy as a function   of $\MMF$ calculated by FCIQMC and  FCIQMC-TC methods for the 54 electron systems with $r_s=0.5$ and 
  $r_s=1.0$.  }
\label{fig_N54}
\end{figure}

Computations on the 54-electron systems are much more expensive and with present computational resources we 
can get converged results only for $r_s=0.5$ and $r_s=1.0$.
For $r_s=0.5$, calculations are performed with 6 different basis sets with $M=246 \rightarrow 1850$, where $N_w=10^8$ is found to be sufficient for all of them.
For $ r_s=1.0$, we can get converged results for $M$ up to 1030. For $M=246, 358$ and $514$, the number of walkers $N_w$ is taken to be $10^{10}$, 
while for $M=778$ and $1030$ we need $N_w=2\times 10^{10}$. 
In Figure \ref{fig_N54},  the results for  54-electron systems are presented as functions of $\MMF$ for $r_s=0.5$ and $r_s=1.0$. The 
asymptotic behaviors are clearer than those of the 14-electron systems for the FCIQMC-TC results and extrapolations are used to evaluate the results in the CBL.

\begin{table}[h]
\caption{Total correlation energies in the complete basis limit for a variety of $N$ and $r_s $ calculated with FCIQMC-TC method.
Extrapolation based on $\MMF$ behavior is used for the 54-electron systems. The results are compared with the previous FCIQMC results \cite{SBA12,NeufeldThom2017}
and the back-flow DMC results\cite{Rios06}.
} 
\label{tab1}
\begin{center}
\begin{tabular}{cc|cccc}
\hline 
$r_s$   &      &   \multicolumn{4} {c} {$E_{\mbox{corr}}$~~(a.u.)   }    \\
\cline{3-6}
$(a.u.)$ &~~N~~  &   ~~ FCIQMC-TC~~ &  ~~FCIQMC\cite{SBA12}~~   & ~~FCIQMC\cite{NeufeldThom2017}~~&   ~~BF-DMC~~    \\  
\hline 
0.5         & 14 &   -0.5948(2)     &  -0.5959(7) &  -0.59467(9) &\\
               &54  &  -2.425(1)        &  -2.435(7)   &  & -2.387(2)  \\   
 1.0        &14  &  -0.5309(2)      &  -0.5316(4) &  -0.5313(2)& \\
              & 54  & -2.134(2)         &  -2.124(3)   &  &-2.125(2) \\
 2.0        &14  &  -0.4440(3)      &  -0.444(1)   &  &\\
 5.0       & 14  &  -0.3078(3)        &  -0.307(1)   &  &\\                    
\hline
\end{tabular}
\end{center}
\end{table}

In Table \ref{tab1}  the complete basis limit   results are presented for all  calculated systems.  The results for the 14-electron systems are simply taken from 
those of  the used largest basis sets. Within the given error bars, these results are already converged to the CBL.  The results for $N=54$ are obtained by 
extrapolations based on the $\MMF$ convergence rate.  
Compared with the previous FCIQMC results, the new results agree  well for the 14-electron systems, with the differences in total energies~$\leq 1~ mE_h$.  
The new results are also in good agreement with the recent high-order Coupled-Cluster study  of the $N=14$ electron system by Neufeld and Thom \cite{NeufeldThom2017}.  
The small differences between the FCIQMC results of Shepherd et al. \cite{SBA12} and Neufeld and Thom \cite{NeufeldThom2017} for the 14-electron systems
arise because of the use of different extrapolation formulae to the infinite-basis set limit, the former being based on $M^{-1}$ whilst the latter 
includes higher order terms  $(b_0 + b_1 M^{-1} + b_2 M^{-2)}$. This allows the use of a larger number of points in the extrapolation procedure (with smaller $M$), 
and leads to somewhat higher extrapolated correlation energies, the implication being that a simple $M^{-1}$ extrapolation tends to over-shoot the exact result unless
a sufficiently large $M$ has been reached. 
Although the differences are not large, this indicates that the precise form of extrapolation
is indeed consequential, and provides a further motivation to try to minimize basis-set errors through analytic means as far as possible.   

For $N=54$, the new transcorrelated result  at  $r_s=0.5$ is $10~mE_h$ above the previous FCIQMC result of Shepherd et al. \cite{SBGA12} based on a $M^{-1}$ extrapolation.
At this density, the FCIQMC-TC result is about $40~mE_h$ below the back-flow DMC result\cite{Rios06} and this indicates that
the fixed-node  error of the BF-DMC result is still quite large for $r_s=0.5$.   For $r_s=1.0$, the previous FCIQMC result for $N=54$ is roughly the same as the
BF-DMC result, while the new result is about $10~mE_h$ below the BF-DMC result. This reveals the fixed nodes error decays very rapidly with $r_s$ 
 and,  we expect that,  at even larger $r_s$ the fixed-node error of BF-DMC will be even smaller and hence can be ignored.

\begin{figure}[h]
\centerline{
\includegraphics[width=0.8\textwidth]{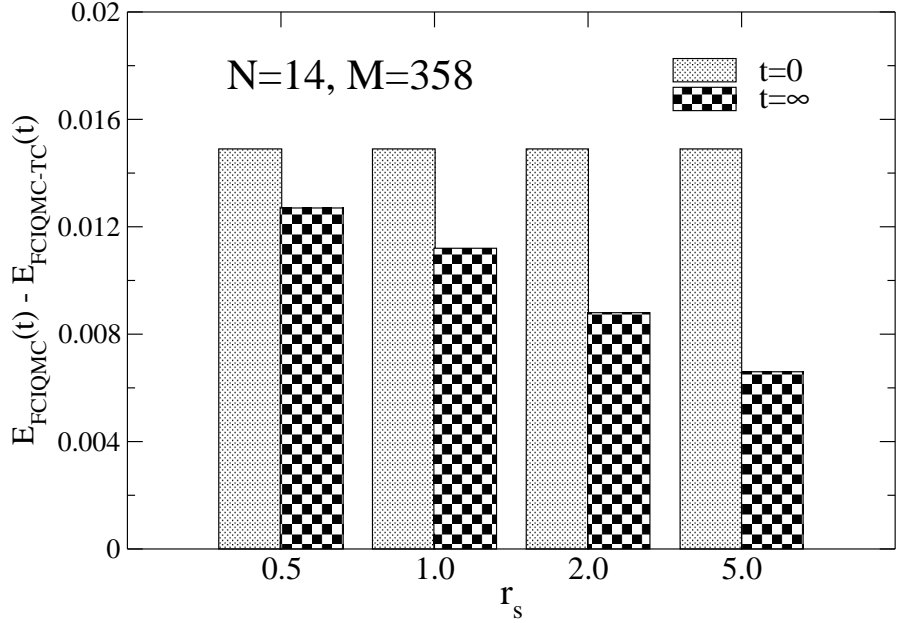}
}
\caption{Energy differences $\Delta E (t)$ at $t=0$ and $t=\infty$ for 14-electron systems with $M=358$ presented for all
different densities $r_s=0.5\sim 5.0$.
}
\label{fig_DE1}
\end{figure}

The effective TC potential in equation (\ref{eq_weff}), now denoting as  
$\tilde{w}_{\eff}({\bf k, p, q})=\tilde{w}({\bf k})+\tilde{w}_{\TC}[\tilde{u}]({\bf k, p, q})$, 
 is constructed with $\tilde{u}({\bf k})$ defined in equation (\ref{eq_uk}).
Since $\tilde{u}({\bf k})$  contains only high frequency terms, it might be expected that the effective potential  $\tilde{w}_{\TC}$ 
has a weak coupling to the low  frequency basis space and thus this potential behaves roughly like a constant potential in
the dynamic evolution. If this would be  the case,  it should be expected that the energy difference produced by  $\tilde{w}_{\TC}$
\begin{equation}
\Delta E(t) = E_{\mbox{FCIQMC}}(t)- E_{\mbox{FCIQMC-TC}}(t),
\end{equation}
should  be approximately independent of imaginary time. In Figure  \ref{fig_DE1},   such energy differences for the 14-electron systems with 
$r_s=0.5\sim 5.0$  and on the basis  $M=358$ is presented for $t=0$ and $t=\infty$. In the calculation, the initial wave function are always 
chosen as the Hartree-Fock (HF) wave function.  It can be seen that at high density  $\Delta E(0)$ is close to
$\Delta E (\infty)$, but at low density they are quite different.  This indicates that the effective potential  $\tilde{w}_{\TC}$ does have a coupling to the 
wave function, especially at low density region.   It turns out that, for a given number of particles $N$ and a given basis size $M$,  $ \Delta E(0)$ is 
independent of $r_s$. This can be easily understood by scaling arguments.  At HF level, the kinetic energy~$\propto  {r_s^{-2}}$ and the exchange
 energy~$\propto r_s^{-1} $,  because the corresponding operators $-\frac 1 2 \sum_i \triangledown_i^2$ and$ \sum_{ij}1/r_{ij}$ scale like 
 $r_s^{-2}$ and $r_s^{-1}$ respectively. By referring to equation \ref{eq_HTC},  we get the expression of $\Delta E (0)$ 
 \begin{equation}
 \Delta E(0)= \langle \Phi_{\mbox{HF}}| \frac 1 2 \sum_i (\triangledown_i \tau)^2|\Phi_{\mbox{HF}} \rangle . 
 \end{equation}
 Based on equation (\ref{eq_tau2}) and equation (\ref{eq_ur}), it is not difficult to find that $ \sum_i (\triangledown_i \tau)^2$ scales as a constant of $r_s$ and,
  therefore $\Delta E(0)$ does not depend on $r_s$.

\begin{figure}[h]
\centerline{
\includegraphics[width=0.8\textwidth]{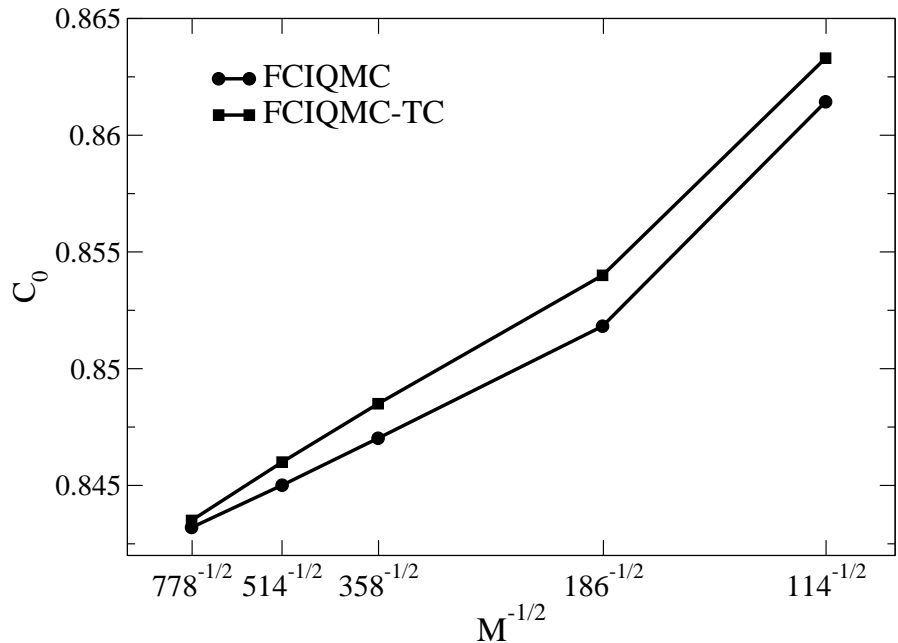}
}
\caption{CI coefficient of the Hartree-Fock determinant in the final solutions of wave function with  the two different methods presented for
different basis sets. The results is calculated for the 14-electron system with $r_s=2.0$.
}
\label{fig_C0}
\end{figure}  
The conclusion that the  effective potential  $\tilde{w}_{\TC}$ couples to the wave function  can also be verified  by looking at the difference of the 
final solutions of wave function between the two different methods. In Figure \ref{fig_C0}, the CI coefficient of the Hartree-Fock determinant ($C_0$)
 in the final solutions of wave function with  the two different methods is  presented for different basis sets. The results is calculated for the 14-electron 
 system with $r_s=2.0$.   $C_0$  of the FCIQMC-TC result  is found clearly larger than that of the FCIQMC result. This  also serves as
 an evidence that the effective potential does couples to the wave function space. In the Figure, it can be found that both curves show a rough $1/\sqrt{M}$
 convergence. For the FCIQMC result, this can be easily understood based on variational argument, that since the energy converges like $\MMO$, the wave
  function should converge like $1/\sqrt{M}$. As for the FCIQMC-TC result, where the energy converges like $\MMF$, the variational argument can not be 
  applied, since the Hamiltonian is not fixed and itself depends on the basis.

\section{Conclusions}  
\label{conclusion}

In this work, we have designed a simple but efficient transcorrelated method for plane wave basis functions. The effective  
Hamiltonian contains only several  two-body operators and thus can be easily implemented.  In order to systematically reduce the error due to the 
neglect of the three-body operator  in the original transcorrelated Hamiltonian,  the correlation factor is constructed in a natural and systematic way 
according to the basis set.  As an initial test, this simple effective Hamiltonian is used in FCIQMC calculations of homogeneous electron gas models. 
The results demonstrate that, with the same computational cost,  this  simple method can improve the FCIQMC convergence rate from $O(\MMO)$ to
 $O\left( \MMF \right)$.  
 
 We have also demonstrated that the effective transcorrelated Hamiltonian does couple to the wave function and changes the dynamic
 evolution of the FCIQMC simulations.  This means that the FCIQMC-TC results can not be precisely estimated based on 
an a posteriori  use of the effective Hamiltonian. 

The effective Hamiltonian,  in principle, can also be applied to other projection methods such as the coupled cluster method. Due to the simple structure 
of the effective Hamiltonian, its implementation should be easier than usual F12 methods.  We are currently working on this implementation. 

Generalization of this method to other type of basis is straightforward and the basis dependent correlation factors can be constructed as follows: take first
 a usual F12 factor, for example a Slater type geminal $f(r_{12})=\exp(-\gamma r_{12})$,  then the correlation factor can be constructed by projection
 against the current (orthonormal) basis set $\{\phi_{i}, \ i=1,\cdots,M  \}$
 \begin{eqnarray}
 u({\bf r}_1, {\bf r}_2)&=&f(|{\bf r}_1- {\bf r}_2|) -\sum_{i, j}^M F_{ij} \phi_i({\bf r}_1) \phi_j({\bf r}_2),\\
   F_{ij}&=& \int  f(|{\bf r}_1- {\bf r}_2|)\phi_i({\bf r}_1) \phi_j({\bf r}_2) d^3 r_1 d^3 r_2 .
 \end{eqnarray}
The effective two-body transcorrelated potential can then be calculated with this correlation factor. Comparing with the plane wave basis, the expressions
may become redundant. Therefore the induced one-body and two-body matrix elements have to be prepared and stored. We plan to work on this in the near future.

\section*{Acknowledgments}
We thank A. Gr\"uneis, J. Brand, P. Jeszenszki and D. Tew  for helpful discussions. Calculations are partly carried out on Hydra clusters in Supercomputing Center  Garching. 

\clearpage

\appendix

\noindent
{\LARGE \bf Appendix}

\section{Regularities and Convergence Rates}
\label{app1}
The regularity of many body wave functions $\Psi$ is $C^{0,1}$ and the non-smoothness comes largely from the cusps between electron pairs.  (In the present analysis we concentrate only on the electronic cusps, and ignore the electron-nuclear cusps.) For a understanding of the relation between the regularity and the convergence rate, it is enough to take the example of a two electron system, which is essentially a one body problem in the center of mass coordinate. Due to the electronic cusp, the short range behavior of the exact wave function for a three dimensional non spin-polarised
system looks like $|\tilde{\psi}({\bf k})| \propto k^{-4}, \ \mbox{for} \ k\rightarrow \infty$. This means that for a finite plane wave basis with a cutoff at $k_c$, the error of wave function due to the missing resolution for the cusp can be estimated as
\begin{equation}
\delta \tilde{\psi} ({\bf k}) \propto \begin{cases}
\frac{1}{k^{4}}, & |{\bf k}|>k_{c},\\
0, & |{\bf k}|\le k_{c}.
\end{cases}
\end {equation}
The error of a variational energy can then be approximated as (ignoring any change of normalisation):  
\begin{eqnarray}
\delta E &\approx \langle  \delta \psi|\hat{H}|\delta \psi \rangle  \nonumber\\
         &\approx& \frac 1 2 \int \delta \tilde{\psi} ({\bf k}) k^2 \delta \tilde{\psi} ({\bf k}) d^3 k \nonumber\\
         &\propto & \frac 1 {k_c^3},
\end{eqnarray}
where we have used that fact that the leading contribution comes from the kinetic energy. By using a Jastrow factor, the regularity of the wave function is improved to $C^{1,1}$,
 which means that now the first order derivatives of $\psi$ is in $C^{0,1}$. This leads to an asymptotic short range behavior of the gradient of the reference function   
$|\widetilde{\triangledown \phi}({\bf k})| \propto k^{-4}, \ \mbox{for} \ k\rightarrow \infty$. Similarly, the energy error for a finite basis set can be estimated as
\begin{eqnarray}
\delta E &\approx& \frac 1 2 \int |\delta \widetilde{\triangledown \phi} ({\bf k})|^2  d^3 k \nonumber\\
         &\propto & \frac 1 {k_c^5}.
\end{eqnarray}
Since the basis size $M\propto k_c^3$, we see that the Jastrow factor improves the convergence rate from ${\cal O}(M^{-1})$ to ${\cal O}\left({\MMF}\right)$.

\bibliographystyle{acs}
\bibliography{/home/hluo/Latex/stybase/quantum}

\end{document}